\documentclass[12pt,a4paper]{article}
\usepackage{psfrag,graphicx}
\usepackage{amsmath}
\usepackage{amsfonts}

\newcommand{\field}[1]{\mathbb{#1}}
\newcommand{\R}{\field{R}}

\newcommand{\thl}{\theta_{\vec{l}} \,}
\newcommand{\thk}{\theta_{\vec{k}} \,}
\def\nb{\nabla^{\bot}}
\def\etaS{\bm{\eta_{S_{\lambda}}}}
\def\xitwo{\xi^2}
\def\g{a}
\def\c{c}
\def\U{U}
\def\Slam{S_{\lambda}}
\def\hlam{h_{\lambda}}
\newcommand\bm[1]{\mbox{\boldmath$#1$}}

\def\Journal#1#2#3#4{{#1} {\bf #2}, #3 (#4)}

\def\JDG{\em J. Diff. Geom.}
\def\CQG{\em Class. Quantum Grav.}

\def\PRD{\em Phys. Rev. D}

\def\JMP{\em J. Math. Phys.}

\def\APP{\em Acta Phys. Polon.}
\def\PRL{\em Phys. Rev. Lett.}

\def\ANYAS{\em Ann. N. Y. Acad. Sci.}

\def\APP{\em Acta Physica Polonica}

\newcommand{\nablab}{D}

\newcommand{\nablanor}{\nabla^{\bot}}

\newtheorem{definition}{Definition}
\newtheorem{lemma}{Lemma}
\newtheorem{theorem}{Theorem}

\begin{document}

\title{Generalized inverse mean curvature flows in spacetime}

\author{Hubert Bray\footnotemark[1] \!, Sean Hayward\footnotemark[2] \! ,
Marc Mars\footnotemark[3] \! and
Walter Simon\footnotemark[4] \\ \\
\footnotemark[1]  Mathematics Department, Duke University, Box 90320  \\
Durham, NC 27708, USA\\ \\ 
\footnotemark[2] Institute for Gravitational Physics and Geometry, \\
The Pennsylvania State University, University Park, PA 16802, USA\\ \\
\footnotemark[3] \! \! \footnotemark[4] Facultad de Ciencias, Universidad de 
Salamanca, \\ Plaza de la Merced s/n, 37008 Salamanca, Spain. \\ \\
\footnotemark[1]  E-mail:  bray@math.duke.edu,
\footnotemark[2] E-mail: sean\_a\_hayward@yahoo.co.uk, \\
\footnotemark[3] E-mail: marc@usal.es, 
\footnotemark[4] E-mail: walter@usal.es}
\maketitle

\begin{abstract}
Motivated by the conjectured Penrose inequality and by the work of Hawking, 
Geroch, Huisken and Ilmanen in the null and the Riemannian case, we examine 
necessary conditions on flows of two-surfaces in spacetime under which the 
Hawking quasilocal mass is monotone. We focus on a subclass of such flows 
which we call uniformly expanding, which can be considered for null as well as 
for spacelike directions. In the null case, local existence of the flow is 
guaranteed. In the spacelike case, the uniformly expanding condition  leaves a 
1-parameter freedom, but for the whole family, the embedding functions satisfy 
a forward-backward parabolic system for which local existence does not hold in 
general. Nevertheless, we have obtained a generalization of the weak 
(distributional) formulation of this class of flows, generalizing the 
corresponding step of Huisken and Ilmanen's  proof of the Riemannian Penrose 
inequality.

\end{abstract}

\section{Introduction}

Penrose has conjectured \cite{Pen} that in an asymptotically flat spacetime 
satisfying the dominant energy condition, the total ADM mass $M$ and the area 
$|S|$ of an outermost apparent horizon $S$  must satisfy
\begin{eqnarray}
\label{PI}
M \geq \sqrt{ \frac{\left |S \right |}{16 \pi}}
\end{eqnarray} 
in units where Newton's gravitational constant is unity. A similar inequality 
should hold for the Bondi mass as well. In the presence of a negative 
cosmological constant $\Lambda$, 
\begin{eqnarray}
M \geq \sqrt{ \frac{\left |S \right |}{16 \pi}} 
\left (\frac{\chi}{2} - \frac{\Lambda}{12 \pi} |S| \right )
\label{PICosm}
\end{eqnarray} 
is conjectured \cite{CS} for suitably defined masses at null infinity, where 
$\chi$ is the Euler characteristic of $S$. Penrose gave a heuristic argument 
for (\ref{PI}) based on the standard view of gravitational collapse, involving 
cosmic censorship, a final stationary configuration and the classification of 
stationary black holes, among others. 

For inequality (\ref{PI}) to hold, it is known that $S$ has to be area outer
minimizing in some sense \cite{GH,KM}. For counterexamples when this condition 
is not fullfilled see \cite{KM,BD}. However, a rigorous proof of this {\it 
Penrose inequality} has been achieved only in two particular cases: in 
spherical symmetry \cite{Hay1,MO,Hay2} and in the time-symmetric case. The 
latter can be formulated as a purely Riemannian problem, termed the {\it 
Riemannian Penrose inequality}, which involves a Riemannian, asymptotically 
Euclidean 3-manifold $(\Sigma,\gamma)$, with non-negative Ricci scalar, which 
translates the dominant energy condition, and with an outermost minimal 
surface $S$, which substitutes the apparent horizon. The inequality (\ref{PI}) 
for this case was proven by Huisken and Ilmanen \cite{HI} for connected $S$ 
and by Bray \cite{B}, with a totally different method, for arbitrary $S$. 
Bray's method uses a flow of Riemannian metrics interpolating between the 
starting metric $\gamma$ and the metric of a $t=\mbox{const.}$ slice of 
Schwarzschild outside the horizon. This flow of metrics keeps the area of the 
outermost minimal surface constant and does not increase the ADM mass. The 
Penrose inequality follows from the fact that equality holds for Schwarzschild.

The approach of Huisken and Ilmanen makes use of the Geroch-Hawking mass
\cite{Ger} which 
is associated to any closed surface $S$ in a hypersurface $\Sigma$:
\begin{eqnarray}
\label{GHMass} 
M_{G}(S) = \sqrt{ \frac{\left | S \right |}{16 \pi}} \left ( 
\frac{\chi(S)}{2} - \frac{1}{16 \pi} \int_S \left( p^2  + \frac{4}{3}\Lambda 
\right) \bm{\eta_S}  \right ),
\end{eqnarray}
where $\bm{\eta_S}$ is the surface element, $|S|$ is the total area of $S$, 
and  $p$ is the mean curvature of $S$ in $\Sigma$. The basic observations due 
to Geroch \cite{Ger} and to Jang and Wald \cite{JW} refer to the case $\Lambda 
= 0$ and are  the following: (\ref{GHMass}) is monotonic for a special flow of 
two-surfaces called the Inverse Mean Curvature Flow (IMCF), approaches the ADM 
mass for suitable surfaces at infinity, such as metric 2-spheres, and equals 
the right hand side of (\ref{PI}) at the minimal surface. Thus, if the IMCF 
exists globally and approaches infinitely large round spheres, the Penrose 
inequality follows. However, the IMCF does in general develop singularities, 
and  Huisken and Ilmanen \cite{HI} succeeded in defining a suitable weak 
version of the flow for which global existence could be proven. This weak 
version of the IMCF allows for jumps and one needs to question monotonicity of 
the Geroch-Hawking mass at the jumps. The condition of connectedness of the 
starting minimal surface was required at this point.

The Geroch-Hawking mass is a translation to 3-dimensional Riemannian manifolds 
of the Hawking mass \cite{Hawking}
\begin{eqnarray}
M_{H}(S) = \sqrt{ \frac{\left | S \right |}{16 \pi}} \left (
\frac{\chi(S)}{2} - \frac{1}{16 \pi} \int_S \left( H^2 + \frac{4}{3}\Lambda
\right) \bm{\eta_S}  \right ),
\label{HawMass}
\end{eqnarray}
defined for 2 surfaces $S$ embedded in spacetime $(V,g)$ with mean curvature 
vector $\vec{H}$, where $H^2 = (\vec{H} \cdot \vec{H})$, the dot denoting the 
inner product on $(V,g)$. If $S$ is embedded in a 3-surface $\Sigma$ such that 
$\vec{H}$ points along the unique unit normal $\vec{\nu}$ of $S$, one can 
write $\vec{H} = p \vec{\nu}$. However, a 2-surface $S$ will in general have 
different Hawking and Geroch-Hawking masses, the former depending only on $S$ 
and the latter also on $\Sigma$. We believe that the Hawking mass is a good 
candidate for proving the general Penrose inequality along the lines sketched 
above for $M_G$ in the Riemannian case. In spherical symmetry, there is a 
quite simple and otherwise general proof based on monotonicity of $M_H$ 
\cite{Hay2b}. To summarize  existing work, to describe the extensions achieved 
in the present paper and to formulate the open issues, we found it useful to 
distinguish the following five steps:

\begin{enumerate}
\item Variation of the Hawking mass
\item Monotonicity properties for special flows
\item Local existence of good flows
\item Global existence of good flows
\item Limits at the horizon and at infinity.
\end{enumerate}

As to the variation of $M_H$, in Sect.\ 2 we find an expression which is valid 
for arbitrary variations and which, when the flow is everywhere spacelike, 
takes the form
\begin{eqnarray}
\frac{d M_{H} (S_{\lambda})}{d \lambda}  =  
 \sqrt{ \frac{ \left | S_{\lambda} \right |}{16 \pi}}  \int_{S_{\lambda}}
\left[\mbox{matter density} +  \mbox{gravitational energy density} +
\mbox{rest}  \right ] \bm{\eta_{S_{\lambda}}}
\label{Var}
\end{eqnarray}
where $\lambda$ denotes the flow parameter and $S_{\lambda}$ the corresponding 
two-surface. The names of the first two terms in the bracket mean that they 
vanish in vacuum and in spherical symmetry, respectively, and they turn out to 
be non-negative under reasonable positive-energy conditions on the matter, and 
under causality conditions on $\vec{H}$ and on the flow vector $\vec{\xi}$. A 
``gravitational energy density'' is not well-defined in general, but the point 
here is to perform some intelligent splitting which allows the  ``rest'' (or 
its integral) to be removed by some simple and not over-restrictive choices of 
the flow.

Such monotonicity conditions are specified in Theorem 1 in Section 3, which is 
the main result of this paper and which we sketch here. To have the energy 
terms non-negative, we impose the dominant energy condition, and we require 
that $\vec{H}$ is spacelike or null with $(\vec{\xi} \cdot \vec{H}) \ge 0$. To 
get rid of the ``rest'' we require two conditions on the flow. The first one 
is the {\it dual inverse mean curvature condition}, reading that $(\vec{\xi} 
\cdot \vec{H}^{\star})$ should be a constant $c(\lambda)$ on each 
$S_{\lambda}$, where $\vec{H}^{\star}$ denotes the dual of $\vec{H}$. For the 
second condition, our splitting suggests two natural alternatives. The first 
one is the {\it inverse mean curvature condition}, namely that $(\vec{\xi} 
\cdot \vec{H}) = a(\lambda)$ with $a(\lambda)$ constant on each $S_{\lambda}$. 
This condition guarantees, in particular, an IMCF on the three-surface spanned 
by the evolving two-surfaces themselves, so it is a natural extension of the 
Geroch condition mentioned above. To get monotonicity of $M_H$, we take the
inverse mean curvature condition and the dual inverse mean curvature condition 
together, for which we adopt the name {\it uniformly expanding} flow. In this 
case, monotonicity still holds if we allow $\vec{\xi}$ to be {\it null} 
somewhere, or everywhere. On regions where either $\vec{\xi}$ or $\vec{H}$ are 
null, the inverse mean curvature condition in fact implies the dual condition, 
and vice versa. If in particular $\vec{\xi}$ is null, this condition has been 
analyzed by Hawking himself \cite{Hawking}, by Eardley \cite{Eardley} and, 
more extensively, by Hayward \cite{Hay1}, while in the spacelike case, the 
uniformly expanding flows were investigated  by Malec, Mars and Simon 
\cite{MMS}. In the special case $c(\lambda) = 0$, the flow is tangent to the 
inverse mean curvature vector and was analyzed by Frauendiener \cite{JF}. 
While $a(\lambda)$ can be set equal to 1 by a reparametrization, if it has no 
zeros,  the presence of  $c(\lambda)$  provides a one-parameter freedom for 
the flows in the spacelike case.

The alternative to the inverse mean curvature condition arising from the
present work is a Poisson equation on $S$ which determines the lapse function 
of the flow, with a source term  depending  on the scalar curvature of $S$ and 
on $H^2$. This equation can always be solved uniquely on any $S_{\lambda}$. 
Here we restrict ourselves to flows which are spacelike everywhere. The 
condition seems new, and has a counterpart in codimension one, with $H^2$ 
replaced by $p^2$, where it can be used as an alternative to IMCF in order to 
obtain monotonicity of the Geroch mass. The properties of this flow will be 
elaborated elsewhere.

The conditions on the flow vector $\vec{\xi}$ as mentioned above guarantee 
monotonicity and they are consistent and solvable on every two-surface 
$S_{\lambda}$. However, this does not mean that we have shown existence of a 
flow in spacetime, i.e.\ a sequence of $S_{\lambda}$, {\it not even locally}. 
For the IMCF in 3-dimensions, the embedding functions satisfy a parabolic 
system for which local existence is guaranteed. In spacetime, the uniformly 
expanding condition gives a local existence result in the case when 
$\vec{\xi}$ is null, since only the null geodesics of the flow have to be 
determined in this case. On the other hand, in the spacelike case we are left 
with a forward-backward parabolic system which was observed by Huisken and 
Ilmanen in the case  $c(\lambda) = 0$ and which we show in general in Sect. 4. 
For such systems, not even local existence results are available.

A key element of the Huisken-Ilmanen proof in codimension 1 is, however, the 
{\it weak} or distributional formulation of the flow in terms of its level 
sets, which satisfy degenerate elliptic equations, if sufficiently 
differentiable. We have obtained a generalization of this variational 
principle to the spacetime case, which we also describe in Sect.\ 4.

We finally comment on point 5 of the list above: $M_H$ obviously gives the 
right hand side of (\ref{PICosm}) on any marginally trapped surface. At 
infinity, provided the $S_{\lambda}$ approach surfaces of constant curvature, 
the Hawking mass approaches expressions which have been studied before: ADM 
and  Bondi for $\Lambda = 0$; for $\Lambda < 0$ see \cite{CS}. Except for the 
asymptotically flat case at spatial infinity, it is an open issue if the flow 
necessarily leads to such surfaces.

\section{Variation of the Hawking mass}

We consider orientable Riemannian manifolds $(V,g)$ and orientable closed (i.e.
compact without boundary) submanifolds $(S,h)$, where $h$ is the induced 
metric on $S$. For our main results, we have to restrict $V$ to be 
4-dimensional with $g$  of signature $(-,+,+,+)$, and $S$ to be  
2-dimensional, with a positive definite metric $h$. We will also consider the 
case that $S$ is embedded in a 3-dimensional manifold $\Sigma$ with positive 
definite metric, which follows from the above one by choosing a suitable 
embedding $\Sigma \subset V$. Some auxiliary results, in particular Lemma 
\ref{normalnormalGauss} and Lemma \ref{arbitraryDim}, hold for arbitrary 
dimensions and arbitrary metric signatures. Up to and including these Lemmas, 
we will consider this general setup, and impose suitable restrictions 
afterwards.


Our aim in this section is to calculate the variation of the Hawking mass 
(\ref{HawMass}) along an arbitrary flow. We take an arbitrary $C^1$ embedding 
\begin{eqnarray*}
\Phi : \hat{S} \times I & \longrightarrow & V, \\ 
(x ,\lambda) & \longrightarrow & \Phi (x,\lambda) \equiv \Phi_{\lambda} (x),
\end{eqnarray*}
where $\hat{S}$ is a copy of $S$ viewed as an abstract manifold detached from 
$V$ and $I \ni 0$ is an open interval of $\R$. The leaves $S_{\lambda} \equiv 
\Phi_{\lambda} (\hat{S})$ are required to be  $C^2$, so that the variation of 
$\vec{H}$ can be defined,
and we write $S$ for $S_{0}$. Choosing an orientation on $(S,h)$ defines 
uniquely a canonical two-form $\bm{\eta_S}$ from the volume form $\bm{\eta}$ 
on $(V,g)$. 

The tangent space $T_p$ of $V$ at $p \in S$ can be decomposed as $T_p = 
T_p^{\Vert} \oplus T_p^{\bot}$ where  $T_p^{\Vert}$ and  $T_p^{\bot}$   are 
the tangent and normal spaces to $S$, respectively, and we shall write 
$\vec{v} = \vec{v}^{\parallel} +\vec{v}^{\bot}$ for any vector $\vec{v} \in 
T_p$. The flow of submanifolds $S_{\lambda}$ has a flow vector $\vec{v} = d 
\Phi (\partial_{\lambda})$ which may have an arbitrary causal character. 
Accordingly, the embedded submanifold ${\cal N} \equiv \Phi (\hat{S} \times I 
)$ need not have constant causal character at different points; in the 
spacetime case, the hypersurface ${\cal N}$ could be timelike, spacelike or 
null at different points. Let us define $\vec{\xi}$ as its normal component 
with respect to $S_{\lambda}$, $\vec{\xi} = \vec{v}^{\bot}$. We shall call 
this vector the {\it normal flow vector} and also simply flow vector when no 
confusion arises.

Evaluating the Hawking mass for any leaf $S_\lambda$, we write $M_H(\lambda) 
\equiv M_H(S_\lambda)$. We will not add a subscript $\lambda$ to $\vec{H}$ or 
to other geometric objects on $\Slam$, unless the meaning is not clear from 
the context.

The derivative of (\ref{HawMass}) with respect to  $\lambda$  involves  
$\pounds_{\vec{\xi}} \, \etaS =  (\vec{\xi} \cdot \vec{H} \,) \etaS$.
The integral of this expression over $S$ yields the first variation of the area.
In terms of the  mean value 
\begin{eqnarray*}
\g(\lambda) \equiv  \frac{\int_{S_{\lambda}}
\left ( \vec{\xi} \cdot \vec{H} \right ) \bm{\eta_{S_{\lambda}}}}{\left | S_{\lambda} \right |},
\end{eqnarray*}
of $(\vec{\xi} \cdot \vec{H})$ over each leaf of the foliation, we can write 
$d |S_{\lambda} |/ d\lambda = \g(\lambda) |S_{\lambda}|$. Using standard 
expressions for derivatives of geometric objects within an integral we now find
\begin{eqnarray}
\label{dM}
\frac{d M_{H} (S_{\lambda})}{d \lambda} = \frac{1}{16 \pi}
\sqrt{ \frac{ \left | S_{\lambda} \right |}{16 \pi}} 
\left  [ 4 \pi \chi(\Slam) \g(\lambda) 
- \int_{\Slam} \left ( 
2 \left ( \pounds_{\vec{\xi}} \vec{H} \cdot \vec{H} \right )
+ (\pounds_{\vec{\xi}} \, g) ( \vec{H}, \vec{H}) +  
\right . \right . \nonumber \\ 
\left  . \left .  + \left [ \left (\vec{H} \cdot \vec{\xi} \right ) + \frac{1}{2} \g(\lambda) \right ] 
\left(H^2 + \frac{4}{3} \Lambda \right) \right ) \etaS \right ].
\end{eqnarray}
Note that only the normal part $\vec{\xi}$ of $\vec{v} = d \Phi 
(\partial_{\lambda})$ appears in this expression because the tangential 
component of $\vec{v}$ gives rise to a divergence of a vector on $T^{\Vert}$ 
which integrates to zero, $S_{\lambda}$ being closed.

In order to reformulate (\ref{dM}), we recall some standard embedding 
formulas, e.g.\ \cite{Jost}, and collect some computations in Lemma 1 and 
Lemma 2 below. Let  $\vec{X}, \vec{Y}$ be tangent vector fields to $S$, 
$\vec{\xi}$ be a normal vector to $S$, and $\nabla$ the covariant derivative 
on $V$. We can decompose the covariant derivatives along $S$ into components 
tangent and normal to $S$ as follows:
\begin{eqnarray}
\label{nabtan}
\nabla_{\vec{X}} \vec{Y} & = & \nablab_{\vec{X}} \vec{Y} - \vec{K} \left ( \vec{X}, \vec{Y} \right ) \\
\label{nabnor}
\nabla_{\vec{X}} \, \vec{\xi} & = & \nabla^{\bot}_{\vec{X}} \, \vec{\xi}  + A_{\vec{\xi}} (\vec{X} ).
\label{Wein}
\end{eqnarray}
Here $\nablab$ is the Levi-Civita connection on $(S,h)$,
$\vec{K}$ is the second fundamental tensor of $S$, 
 $\nablanor$ is the connection on the normal bundle of $S$ 
and $A_{\vec{\xi}} : T_p S \rightarrow T_p S$ is the Weingarten map, both 
defined by the decomposition in (\ref{Wein}).

\begin{lemma}
\label{normalnormalGauss}
Let $S$ be a $C^2$  embedded submanifold of arbitrary codimension in a Riemannian
manifold $(V,g)$ of arbitrary signature and assume that $S$ has a
non-degenerate induced metric.
Let $\vec{X},\vec{Y}$ be vector fields tangent to $S$ and
$\vec{\xi}$ a $C^2$ vector field normal to $S$. Then
\begin{eqnarray*}
\left[(\pounds_{\vec{\xi}}\vec{K})(\vec{X},\vec{Y}) \right]^{\bot} = 
 - \nb_{\vec{X}} \nb_{\vec{Y}} \, \vec{\xi} + \nb_{\nablab_{\vec{X}} \vec{Y}} \,
\vec{\xi}
+ \vec{K} \left ( A_{\vec{\xi}} ( \vec{Y} ), \vec{X} \right ) \\ 
-  \left ( \nabla_{\vec{K} (\vec{X},\vec{Y} )} \vec{\xi} \right )^{\bot}
 - \left ( R(\vec{\xi}, \vec{X} ) \vec{Y} \right )^{\bot}. 
\end{eqnarray*}
\end{lemma}
{\it Proof}. By construction $d \Phi_{\lambda} (\vec{X})$ is a vector field 
tangent to $\Slam$. It follows directly that the commutation of the velocity 
vector $\vec{v} = d \Phi (\partial_{\lambda})$ and $\vec{X}$ gives a vector 
field tangent to $S$ and the same happens with $[\vec{\xi}, \vec{X} ]$, i.e. 
$[ \vec{\xi}, \vec{X} ]^{\bot} = 0$. 
Using  (\ref{nabtan}) and (\ref{nabnor}) and the definition 
$R(\vec{\alpha},\vec{\beta}) \vec{\gamma} = \nabla_{\vec{\alpha}} 
\nabla_{\vec{\beta}} \vec{\gamma} - \nabla_{\vec{\beta}} \nabla_{\vec{\alpha}} 
\vec{\gamma} - \nabla_{[\vec{\alpha},\vec{\beta}]} \vec{\gamma}$ for the 
Riemann tensor we get
\begin{eqnarray*}
\left ( R(\vec{\xi}, \vec{X} ) \vec{Y} \right )^{\bot} & = &
\left( \nabla_{\vec{\xi}} \nabla_{\vec{X}} \vec{Y}  - 
\nabla_{\vec{X}} \nabla_{\vec{\xi}} \vec{Y} \right)^{\bot} 
+ \vec{K} \left ( \vec{Y}, [\vec{\xi}, \vec{X}] \right) 
= \\
{} & = & 
\left(\nabla_{\vec{\xi}} D_{\vec{X}} \vec{Y}  - 
\nabla_{\vec{\xi}}  \vec{K}(\vec{X},\vec{Y}) -
\nabla_{\vec{X}} \nabla_{\vec{Y}} \vec{\xi}  - 
 \nabla_{\vec{X}} \, [\vec{\xi}, \vec{Y}] \right )^{\bot}
 + \vec{K} \left ( \vec{Y}, [\vec{\xi}, \vec{X}] \right) 
= \\
{} & = & \nb_{D_{\vec{X}} \vec{Y}} \, \vec{\xi} -  \left(\nabla_{\vec{\xi}}  \vec{K}
(\vec{X},\vec{Y})\right)^{\bot} -
\nb_{\vec{X}} \nb_{\vec{Y}} \, \vec{\xi} +  \vec{K} \left ( A_{\vec{\xi}} \, (\vec{Y} ), \vec{X} \right )
+ \\ 
{} & + & \vec{K} \left ( \vec{X}, [\vec{\xi}, \vec{Y} ]  \right ) 
+ \vec{K} \left ( \vec{Y}, [\vec{\xi}, \vec{X} ]  \right ).
\end{eqnarray*}
Using now $\nabla_{\vec{\xi}} \, ( \vec{K}(\vec{X}, \vec{Y} )) =
\nabla_{\vec{K}(\vec{X},\vec{Y} )} \vec{\xi} + \pounds_{\vec{\xi}} ( \vec{K} (\vec{X},\vec{Y}) )$
and the fact that $\pounds_{\vec{\xi}} \, ( \vec{K} (\vec{X},\vec{Y} )) =
\pounds_{\vec{\xi}} ( \vec{K} ) (\vec{X},\vec{Y} ) + \vec{K} ( [\vec{\xi}, \vec{X}], \vec{Y} )
+ \vec{K} ( \vec{X}, [\vec{\xi}, \vec{Y}])$, the Lemma follows directly. \hfill $\Box$

We now treat the non-trivial term $ \pounds_{\vec{\xi}} \vec{H}$ in (\ref{dM}) 
as follows. As $\vec{H}$ is normal to $S$, only $(\pounds_{\vec{\xi}} 
\vec{H})^{\bot}$ needs to be calculated.
We choose a basis $\{ \vec{e}_A \}$ of the tangent space $T_p^{\Vert}$. 
Without loss of generality, we can assume that this basis is Lie propagated 
along the variation vector $\vec{\xi}$, i.e. $[\vec{\xi},  \vec{e}_A ]=0$. 
Indices $A,B,\cdots$ therefore refer to tensor objects within $S$ expressed on 
this basis. For instance $\vec{K}_{AB} = \vec{K} (\vec{e}_A, \vec{e}_B)$, and 
$\nabla_A =  \nabla_{\vec{e}_A}$. Such indices are lowered and raised with the 
induced metric $h_{AB}$ and its inverse $h^{AB}$. Directly from the definition 
of 
\begin{equation}
\vec{H}=\mbox{tr}_{\Slam}\vec{K}=h^{AB}\vec{K}_{AB} 
\end{equation}
and $\partial_{\lambda} h_{\lambda}^{AB} = - 2 ( K_{\lambda}^{AB} \cdot 
\vec{\xi} \, )$, we get, after making use of Lemma \ref{normalnormalGauss},
\begin{eqnarray}
\left ( \pounds_{\vec{\xi}} \vec{H}_{\lambda} \right )^{\bot}  = 
- 2 \vec{K}_{AB} \left ( \vec{K}^{AB} \cdot \vec{\xi} \right) - h^{AB}\nb_{A} \nb_{B}
\vec{\xi} + h^{AB} \nb_{\nablab_A \vec{e}_B} \vec{\xi}  \nonumber \\
+ h^{AB} \vec{K} \left ( A_{\vec{\xi}} \, ( \vec{e}_A),\vec{e}_B \right ) 
 - h^{AB} \left [ R ( \vec{\xi}, \vec{e}_A ) \vec{e}_B \right ]^{\bot} -
\left ( \nabla_{\vec{H}\, } \vec{\xi} \, \right )^{\bot}. 
\label{dHnor}
\end{eqnarray}
The first and fourth terms are the same apart from coefficient, while the 
second and third terms combine as 
$-\mbox{tr}_{\Slam}\left(\nb\nb\vec{\xi}\right)$. Also, inserting 
(\ref{dHnor}) in (\ref{dM}) we observe that the last term, in 
$\nabla_{\vec{H}} \vec{\xi}$, cancels with $(\pounds_{\vec{\xi}} \, g) ( 
\vec{H}, \vec{H} )$. Notice that this term is not intrinsic for the variation 
because it depends on how $\vec{\xi}$ changes along normal directions. It is 
clear that such dependence cannot occur in a first derivative. Summarizing, we 
have the following result.
\begin{lemma}
\label{arbitraryDim} With the same assumptions as in Lemma 
\ref{normalnormalGauss}, suppose also that $S$ is closed and orientable. Then, 
the variation of the Hawking mass along $\vec{\xi}$ reads
\begin{eqnarray}
\frac{d M_{H} (S_{\lambda})}{d \lambda} & = & \frac{1}{16 \pi} \sqrt{ \frac{ \left | S_{\lambda} \right |}{16 \pi}} 
\left[ \frac{}{} 4 \pi \chi(\Slam) \g(\lambda) \right. + \nonumber\\
{} & + & \left.  \int_{\Slam} \left( 2  ( \vec{K}_{AB} \cdot \vec{H} )
( \vec{K}^{AB} \cdot \vec{\xi}  )   
- \left [ \left(\vec{\xi} \cdot \vec{H} \right) + \frac{1}{2} \g(\lambda) \right]
\left(H^2 + \frac{4}{3}\Lambda \right)  \right. \right. + \nonumber\\
 & + & \left. \left. 2 h^{AB} \left ( R (\vec{\xi}, \vec{e}_A) \vec{e}_B \cdot \vec{H} \right ) 
+ 2 \mbox{tr}_{\Slam} \left ( \vec{H} \cdot \nb \nb \vec{\xi} \right ) \right ) \etaS \right]  
\label{dMhGen}
\end{eqnarray}
\end{lemma}

We now restrict ourselves to spacelike two-dimensional surfaces in a four 
dimensional spacetime of Lorentzian signature. In particular, the normal 
spaces $T^{\bot}_{\lambda}$ are now two-dimensional and Lorentzian, and 
inherit an orientation from the orientation of $S$. We can introduce 
projection operators $P^{\mu\nu}_{\Vert}$ and $P^{\mu\nu}_{\bot}$ which act as 
the identities on $T_p^{\Vert}$ and $T_p^{\bot}$, respectively, and annihilate 
vectors in the other orthogonal spaces. In particular, we have $g^{\mu\nu} = 
P^{\mu\nu}_{\bot} + P^{\mu\nu}_{\Vert}$. 

This setup also allows us to define the Hodge dual operation on normal fields 
$\vec{W}$, \begin{eqnarray*} {W^{\star}}^{\alpha} = \frac{1}{2} 
\eta^{\alpha}_{\,\,\,\beta\gamma\delta} W^{\beta} e^{\gamma}_A e^{\delta}_B 
\eta_{\Slam}^{AB}, \end{eqnarray*} We have $( \vec{W} ^{\star \star} ) = 
\vec{W}$ and $( \vec{W} \cdot \vec{U}^{\star} ) = - ( \vec{W}^{\star} \cdot 
\vec{U} )$. Therefore any normal vector is orthogonal to its dual and both 
have opposite norms. In particular, a normal null vector is either self-dual, 
$\vec{l}^{\star} = \vec{l}$, or anti-self dual, $\vec{l}^{\star} = -\vec{l}$. 
We note the following completeness property, valid for any pair of vectors 
$\vec{u}, \vec{v} \in T^{\bot}_p $: 

\begin{equation} 
\label{comp} v^{\alpha} u^{\beta}+ u^{\alpha} v^{\beta} - v^{\star \alpha} 
u^{\star \beta}- u^{\star \alpha} v^{\star \beta} = 2 (\vec{v}\cdot\vec{u}) 
P^{\alpha\beta}_{\bot}.
\end{equation} 
This expression can be proven directly by noting that both terms are 
symmetric, orthogonal to $S_{\lambda}$ on each index and give an identity when 
contracted with any tensor product of   $u_{\alpha}, v_{\alpha}, 
u_{\alpha}^{\star}, v_{\alpha}^{\star}$. In particular, when $\vec{k}$ and 
$\vec{l}$ are null and linearly independent, with $\phi  \equiv - (\vec{l} 
\cdot \vec{k} ) \neq 0$, we have 

\begin{equation} 
\label{compnull}
k^{\alpha} l^{\beta}+ l^{\alpha} k^{\beta} = - \phi P^{\alpha\beta}_{\bot}.
\end{equation} 
We next introduce the Ricci scalar  $R(h_{\lambda})$ of  $(\Slam,h_{\lambda})$,
and the trace-free part $\vec{\Pi} = \vec{K} - \frac{1}{2} \vec{H} h$ of the 
extrinsic curvature vector $\vec{K}$.Furthermore, we define 
$\Theta^T_{\mu\nu}$, the ``transverse part of the gravitational energy'' by
\begin{eqnarray}
\label{TTh} 8 \pi \Theta^T(\vec{H}^{\star},\vec{\xi}^{\star}) & = & 
\left(\vec{\Pi}_{AB}\cdot\vec{H}\right)\left(\vec{\xi} \cdot \vec{\Pi}^{AB} 
\,\right) 
-\frac12\left(\vec{\Pi}_{AB}\cdot\vec{\Pi}^{AB}\right)\left(\vec{\xi} \cdot 
\vec{H}\, \right)
 \nonumber \\
{}& = & \left(\vec{\Pi}_{AB}\cdot\vec{H}^\star\right)\left(\vec{\xi}^\star 
\cdot \vec{\Pi}^{AB} \, \right) - 
\frac12\left(\vec{\Pi}_{AB}\cdot\vec{\Pi}^{AB}\right)\left(\vec{\xi}^\star 
\cdot \vec{H}^\star \,\right)
\end{eqnarray}
for any pair of vectors $\vec{H},\vec{\xi}$.
This terminology will be justified later, and the equivalence of the two
alternative expressions in (\ref{TTh}) follows from (\ref{comp})
with $\vec{v} \rightarrow \vec{H}$ and $\vec{u} \rightarrow \vec{\xi}$. 
We also define
\begin{equation}
\label{Om}
\Omega = \frac{1}{2} R(h_{\lambda}) -  \frac{1}{4} H^2 - \frac{1}{3} \Lambda 
\end{equation}
which may be called the ``Hawking energy density'' since, using
the Gauss-Bonnet formula, 
\begin{equation}
\label{GB}
 \int_{\Slam} R (h_{\lambda} ) \etaS = 4 \pi \chi (\Slam)
\end{equation}
we find from (\ref{HawMass}) that
\begin{equation}
\label{Haw}
M_{H}(S) = \sqrt{ \frac{\left | S \right |}{16 \pi}} \left (
 \frac{1}{4 \pi} \int_S \Omega \bm{\eta_S}  \right ).
\end{equation}

\begin{lemma}
\label{derMH1} 
Let $(V,g)$ be an oriented 4-dimensional spacetime of signature 
$(-,+,+,+)$ and $S$ a 2-dimensional oriented closed and spacelike $C^2$ 
surface in $V$. Let $\Phi$ denote a flow of two-surfaces as defined above. 
Then, the derivative of the Hawking mass along the flow reads
\begin{eqnarray}
\frac{d M_{H} (S_{\lambda})}{d \lambda}  & = & 
\frac{1}{8 \pi} \sqrt{ \frac{ \left | S_{\lambda} \right |}{16 \pi}}  \int_{S_{\lambda}}
\left[ \left ( G (\vec{H}^{\star}, \vec{\xi}^{\star} ) + \Lambda  (\vec{H}^{\star} \cdot \vec{\xi}^{\star} \,  )
\right )
 +  
8 \pi \Theta^T (\vec{H}^{\star},\vec{\xi}^{\star} \, ) + \right. {}  \nonumber \\
 & + & \left.
\mbox{tr}_{S_{\lambda}} \left ( \vec{H} \cdot \nablanor \nablanor \vec{\xi} \right )
-  \Omega \left[ \left (\vec{\xi} \cdot \vec{H} \right ) - \g(\lambda) \right ] \right ] 
\bm{\eta_{S_{\lambda}}},
\label{dMH1}
\end{eqnarray}
where $G$ is the Einstein tensor of $(V,g)$.
\end{lemma}
{\it Proof}. 
We first note that
\begin{eqnarray}
\label{Ricc}
R_{\alpha\beta} H^{\alpha} \xi^{\beta} = R_{\alpha\beta} H^{\star \, \alpha} \xi^{\star \, \beta} 
+ \left ( \vec{\xi} \cdot \vec{H} \right ) P^{\alpha\beta}_{\bot}
R_{\alpha\beta}.
\end{eqnarray}
which follows from (\ref{comp}) with $\vec{v} \rightarrow \vec{H}$ and 
$\vec{u} \rightarrow \vec{\xi}$ by contracting with $R_{\alpha\beta}$. We next 
recall the Gauss identity
\begin{eqnarray}
\label{Gauss} 
P^{\alpha\beta}_{\Vert} P^{\gamma\delta}_{\Vert} 
R_{\alpha\gamma\beta\delta} = R (h_{\lambda}) - H^2 + \left ( \vec{K}_{AB} \cdot \vec{K}^{AB} \right ).
\end{eqnarray}
We can now show that  
\begin{eqnarray}
\label{Riem}
 h^{AB} \left ( R (\vec{\xi}, \vec{e}_A) \vec{e}_B \cdot \vec{H} \right ) = 
G (\vec{H}^{\star}, \vec{\xi}^{\star} \, ) + \frac{1}{2} \left (\vec{\xi} 
\cdot \vec{H} \right ) \left  [ H^2 - R(h_{\lambda}) - \left ( \vec{K}_{AB} 
\cdot \vec{K}^{AB} \right ) \right ]
\end{eqnarray}
by decomposing $\vec{\xi}$, $\vec{H}$ and $P_{\bot}^{\mu\nu}$ in a null basis 
$\vec{k}, \vec{l}$, using (\ref{compnull}) for $P_{\bot}^{\mu\nu}$ and working 
out the contractions, using also (\ref{Ricc}) and (\ref{Gauss}).  
Finally, we obtain (\ref{dMH1}) by inserting (\ref{Riem}) into (\ref{dMhGen}) 
and making use of the Gauss-Bonnet formula (\ref{GB}). \hfill $\Box$

We now reformulate (\ref{dMH1}) in the case where $\vec{\xi}$ is not null, and 
we write $(\vec{\xi} \cdot \vec{\xi} \, )  = \xi^2 = \epsilon e^{2\psi}$ for 
$\epsilon =\pm1$ and some function $\psi$. We decompose $\vec{\xi}$ in a null 
basis as follows:
\begin{equation}
\label{xis}
\vec{\xi} = \vec{\xi}_l+\vec{\xi}_k \qquad \mbox{where}~~\vec{\xi}_l = 
A\vec{l},~~~ \vec{\xi}_k = B\vec{k}~~ \mbox{for some functions}~~ A, B.
\end{equation}
Then $\epsilon e^{2\psi} = -2AB\phi \neq 0$. 
Furthermore, we introduce the following one-form on $\Slam$
\begin{eqnarray*}
U_A = \frac{1}{2\phi}\left(\frac{\vec{l}\cdot\nabla_{A} 
\vec{\xi}_k}{B}-\frac{\vec{k}\cdot\nabla_{A} \vec{\xi}_l}{A}\right),
\end{eqnarray*}
and the ``longitudinal part of the gravitational energy density'' 
$\Theta^{L}(\vec{\xi}\,)_{\mu\nu}$
\begin{eqnarray}
\label{TL}
8 \pi \Theta^L(\vec{\xi}\,) (\vec{H}^{\star},\vec{\xi}^{\star}\,) =(|U|^2+|D\psi|^2)(\vec{\xi}\cdot\vec{H}) -(2U\cdot
D\psi)(\vec{\xi} \cdot\vec{H}^\star).
\end{eqnarray}
\begin{lemma}
\label{derMH2} We require the assumptions of Lemma \ref{derMH1} and in addition
that $\vec{\xi}$ is nowhere null. We obtain
\begin{eqnarray}
\frac{d M_{H} (S_{\lambda})}{d \lambda}  = 
\frac{1}{8 \pi} \sqrt{ \frac{ \left | S_{\lambda} \right |}{16 \pi}}  \int_{S_{\lambda}}
\left[ \left ( G (\vec{H}^{\star},\vec{\xi}^{\star} \, )  + \Lambda ( \vec{H}^{\star} \cdot \vec{\xi}^{\star} ) 
\right )
+ 8 \pi \Theta^T (\vec{H}^{\star},\vec{\xi}^{\star} \,) + 
\right. {} \hspace{1cm} \nonumber\\ 
\left .  +  8 \pi \Theta^L(\vec{\xi} \,) (
\vec{H}^{\star},\vec{\xi}^{\star} \, )  - 
D.U (\vec{\xi} \cdot \vec{H}^{\star}) + 
\left[ \left (\vec{\xi} \cdot \vec{H} \right ) - \g(\lambda) \right ]
\left ( \Delta \psi - \Omega \right )  \right ] \bm{\eta_{S_{\lambda}}}
\label{dMH2}
\end{eqnarray}
where $D.U$ denotes the divergence of $U$.
\end{lemma}
{\it Proof.} In terms of the basis introduced above, we have
\begin{eqnarray*} 
\nb\vec{\xi} = (D\psi+U)\vec{\xi}_l + (D\psi-U)\vec{\xi}_k.
\end{eqnarray*}
It follows that
\begin{eqnarray*} 
\mbox{tr}_{\Slam}\left(\nablanor\nablanor\vec{\xi}\,\right) = 
\left(\Delta\psi+D\cdot U+|D\psi+U|^2\right)\vec{\xi}_l +  \left(\Delta \psi-D\cdot 
U+|D\psi-U|^2\right)\vec{\xi}_k
\end{eqnarray*}
and hence
\begin{eqnarray} 
\mbox{tr}_{\Slam}\left(\vec{H}\cdot\nablanor\nablanor\vec{\xi}\,\right) = 
\left(\Delta \psi+|D\psi|^2+|U|^2\right)(\vec{\xi}\cdot\vec{H}) - \left(D\cdot 
U+2U\cdot D\psi\right)(\vec{\xi}\cdot\vec{H}^{\star}). \label{Derderxi} %
\end{eqnarray}
Inserting this in (\ref{dMH1}) and using (\ref{TL}) gives (\ref{dMH2}). \hfill $\Box$

\section{Monotonicity properties of the Hawking mass}

Lemmas \ref{derMH1} and  \ref{derMH2} are the basis for a systematic 
discussion of monotonicity of the Hawking mass. In both expressions there are 
four types of terms. We shall discuss them separately and show that each of 
them is non-negative under suitable conditions on the flow. Finally we put 
these conditions together to obtain the  monotonicity result for $M_H$, 
Theorem 1. 

As to the integral 
\begin{eqnarray}
\label{IG} \int_{S_{\lambda}}\left[  G (\vec{H}^{\star},\vec{\xi}^{\star}\,) + 
\Lambda \left (H^{\star} \cdot\vec{\xi}^{\star} \right )\right] 
\bm{\eta_{S_{\lambda}}}
\end{eqnarray}
we require that
\begin{equation}
\label{ec}
(G_{\alpha\beta} + \Lambda g_{\alpha\beta}) u^{\alpha} v^{\beta} \geq 0 
\end{equation}
for any pair of future (or past) causal vectors. If the Einstein field 
equations hold, the bracket is equal to $8\pi$ times the energy momentum 
tensor. In any case, we call (\ref{ec})  the dominant energy condition. 
Accordingly, $(\ref{IG})$ is non-negative if $\vec{H}^{\star}$ and 
$\vec{\xi}^{\star}$ are future (past) causal.

In order to consider the following terms in (\ref{dMH1}), it is convenient to 
define four closed subsets of the normal space $T_p^{\bot}$. Being a 
Lorentzian vector space, we can speak of causal (i.e.\ timelike or null) 
vectors and of achronal (i.e.\ spacelike or null) vectors. Let us denote by 
$C^{\pm}_p$ the subset of future (past) causal vectors and $A^{\pm}_p$ the set 
of vectors such that its Hodge dual is future (past) causal. Notice that all 
the vectors in $A^{\pm}_p$ are achronal. These four sets clearly cover the 
normal space $T_p^{\bot}$ and they are not disjoint. The $\vec{0}$ vector 
belongs to all of them. Furthermore, the intersection $C^{+}  \cap A^{+}$, 
dropping the subindex $p$ from now on for simplicity, is the set of self-dual 
future null vectors, and similarly for $C^{-} \cap A^{-}$. As for $C^{+} \cap 
A^{-}$, this is the set of anti-self dual future null vectors, and past for 
$C^{-} \cap A^{+}$. This decomposition is visualized in Figure 1.

\begin{figure}
\begin{center}
\begin{psfrags}
\includegraphics[angle=0,totalheight=6.5cm]{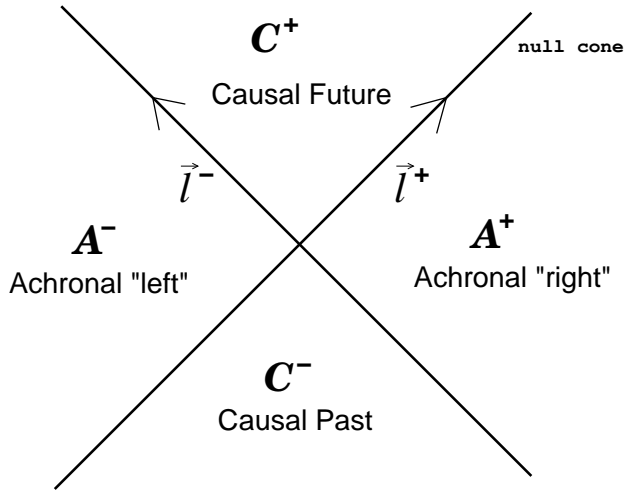}
\end{psfrags}
\caption{Decomposition of the normal space into four
closed sets.}
\label{dhf}
\end{center}
\end{figure}
A positivity Lemma for $\Theta^T (\vec{H}^{\star},\vec{\xi}^{\star}\,)$ 
follows directly from the positivity result for the so-called super-energy 
tensors, Theorem 4.1 in \cite{SE}. Since our case only requires a simpler 
version of the result, we add a proof for completeness.
\begin{lemma}
\label{LT}
The inequality
\begin{eqnarray}
2 \left ( \vec{\Gamma}_{AB} \cdot \vec{H} \right ) \left (
\vec{\Gamma}^{AB} \cdot \vec{\xi} \, \right ) - 
\left ( \vec{\Gamma}_{AB} \cdot \vec{\Gamma}^{AB} \right )
\left (\vec{\xi} \cdot  \vec{H}\, \right ) \geq 0
\label{inequality1}
\end{eqnarray}
holds for {\mbox any} tensor $\vec{\Gamma}_{AB}$ if and only 
if $\vec{\xi}, \vec{H} \in C^{+}$, $\vec{\xi}, \vec{H} \in C^{-}$, $\vec{\xi}, 
\vec{H} \in A^{+}$ or $\vec{\xi}, \vec{H} \in A^{-}$. 
\end{lemma}
{\it Proof:} The ``only if'' part of the Lemma is easy to prove by finding 
suitable counterexamples. For the direct part, let $p$ be any point in $S$. 
The object  $2 \Gamma_{AB \,\mu} \Gamma^{AB}_{\nu} - \Gamma_{AB}^{\rho} 
\Gamma^{AB}_{\rho} \gamma_{\mu\nu} |_p $ defines a map on $T_p N \times T_p N$ 
which is obviously continuous. Thus, it suffices to prove the inequality 
almost everywhere on the subspace $(C^{+} \times C^{+}) \cup (C^{-} \times 
C^{-}) \cup (A^{+} \times A^{+}) \cup (A^{-} \times A^{-} )$. We can assume 
without loss of generality that $\vec{H}$ and $\vec{\xi}$ are both non-null 
and linearly independent. Since, by assumption, both vectors belong to the 
same subset $C^{+}$, $C^{-}$, $A^{+}$ or $A^{-}$, we have  $H^2 = \epsilon 
a^2$, $\xi^2 = \epsilon b^2$, where $\epsilon = \pm 1$ and $a,b$ are strictly 
positive. Moreover we have $\epsilon (\vec{\xi} \cdot \vec{H} ) > 0$. On any 
two-dimensional Lorentzian space, the inequality $(\vec{u} \cdot \vec{v} )^2 
\geq v^2 u^2$ holds for any pair of vectors. Hence $\epsilon (\vec{\xi} \cdot 
\vec{H} ) \geq ab$. Since $\vec{H}$ and $\vec{\xi}$ are linearly independent, 
we can decompose $\vec{\Gamma}_{AB} = \Sigma_{AB} \vec{H} + \Omega_{AB} 
\vec{\xi}$. Direct substitution into the left hand side of 
({\ref{inequality1}) gives
\begin{eqnarray*}
\epsilon \left ( \vec{\xi} \cdot \vec{H} \right ) \left ( \Sigma_{AB} \Sigma^{AB} a^2 + \Omega_{AB} 
\Omega^{AB} b^2 \right ) + 2 a^2 b^2 \Sigma_{AB} \Omega^{AB} \geq ab \left ( a^2 \Sigma_{AB} \Sigma^{AB}
\right . \\ \left . 
+ b^2 \Omega_{AB} \Omega^{AB} + 2 ab \Sigma_{AB} \Omega^{AB} \right ) = a b 
\left ( a \Sigma_{AB} + b \Omega_{AB} \right ) \left ( a \Sigma^{AB} + b \Omega^{AB} \right ) \geq 0 
\end{eqnarray*}
and the Lemma is proven. \hfill $\Box$
 
Putting $\Gamma_{AB} = \Pi_{AB}$, it follows that $\Theta^T 
(\vec{H}^{\star},\vec{\xi}^{\star}\,)$ is monotone iff the velocity vector of 
the flow points into the same quadrant as the mean curvature vector, at each 
point on the surface, or equivalently if $\vec{\xi}, \vec{H} \in C^{+}$, 
$\vec{\xi}, \vec{H} \in C^{-}$, $\vec{\xi}, \vec{H} \in A^{+}$ or $\vec{\xi}, 
\vec{H} \in A^{-}$.

We turn now to the positivity of
 $\Theta^L(\vec{\xi}) (\vec{H}^{\star},\vec{\xi}^{\star} )$. This object is defined
only when $\vec{\xi}$ is non-null. However, equation (\ref{TLpos}}) will be 
used later in the limit when $\vec{\xi}$ tends to a null vector.
\begin{lemma}
\label{ThetaL} Let $\vec{\xi}$ be non-null everywhere on $\Slam$ and asume 
that either (i) $\vec{\xi} \in C^{+}, \vec{H} \in C^{-}$, (ii) $\vec{\xi} \in 
C^{-}, \vec{H} \in C^{+}$, (iii) $\vec{\xi}, \vec{H} \in A^{+}$, 
\underline{or}  (iv) $\vec{\xi}, \vec{H} \in A^{-}$ holds everywhere on 
$S_{\lambda}$. Then $\Theta^L(\vec{\xi}) (\vec{H}^{\star},\vec{\xi}^{\star} ) 
\geq 0$.

\end{lemma}

{\it Proof:} Conditions (i), (ii), (iii) or (iv) together with the fact that
$\vec{\xi}$ is not null implies $(\vec{\xi} \cdot \vec{H}) \neq 0$ everywhere. We
rewrite (\ref{TL}) as
\begin{eqnarray}
\label{TLpos}
8 \pi \Theta^L(\vec{\xi}\,) (\vec{H}^{\star},\vec{\xi}^{\star}\,) = \left(\vec{\xi}\cdot\vec{H}\right)\left|U-\frac{(\vec{\xi}\cdot 
\vec{H}^{\star})}{(\vec{\xi}\cdot\vec{H})}D\psi\right|^2 
+\frac{H^2\xi^2}{(\vec{\xi}\cdot\vec{H})}|D\psi|^2 
\end{eqnarray}
after using  $H^2 \xi^2 = ( \vec{\xi} \cdot \vec{H} )^2
 -  ( \vec{\xi} \cdot \vec{H}^{\star} ) ^2$, which follows from (\ref{comp}). Each of the
hypotheses (i) to (iv) implies $H^2 \xi^2 \geq 0$ and $(\vec{H} \cdot 
\vec{\xi} ) >0$, and the Lemma follows. \hfill $\Box$.

Turning now  to the other terms in the derivative of the Hawking mass 
(\ref{dMH1}), we first assume that  that $\vec{\xi}$ is non-null everywhere, 
in which case we can use (\ref{derMH2}).  The last term, namely
\begin{eqnarray}
\label{IP}
\int_{S_{\lambda}}  \left[ \left ( \vec{\xi} \cdot \vec{H} \right ) - \g(\lambda) \right ]
\left ( \Delta \psi - \Omega \right )  \bm{\eta_{S_{\lambda}}},
\end{eqnarray}
involves the product of two factors. We first note  that the geometric flow of 
2-surfaces is obviously independent of the parametrization used, and a 
reparametrization changes the velocity field $\vec{\xi}$ by a nowhere-zero 
constant on each leaf. In particular, such a reparametrization rescales 
$\g(\lambda)$ by an arbitrary nowhere-zero function of $\lambda$. Thus, if 
$\g(\lambda) \neq 0$ everywhere, $\g=1$ can be assumed without loss of 
generality. If, on the other hand, $\g$ has zeros, be it at isolated points or 
on intervals, those values are meaningful and independent of reparametrization.

The first factor in (\ref{IP}) is the difference between 
$(\vec{\xi} \cdot \vec{H} \,)$ and its mean value on $S_{\lambda}$, 
$\g(\lambda)$. Consequently, it will  never have a constant sign on the 
surface unless it is identically zero.  While it might be possible to define a 
flow in such a way that each factor changes sign at the same place, such a 
condition would be very complicated to analyze. Hence we restrict ourselves to 
the cases that either the first factor is identically zero or that the second one is constant,
each of which remove the integral (\ref{IP}).

As to the first option, the condition $(\vec{\xi}, \vec{H})= \g(\lambda)$  is 
the generalization to codimension two of the IMCF on surfaces in a 
three-dimensional space. Indeed, in codimension one,  the mean curvature 
vector and the normal velocity are obviously parallel, and setting $a=1$,
assuming $a(\lambda) \ne 0$, this condition just states $\vec{\xi} = \vec{\mu} 
/p$, where $\vec{\mu}$ is the unit normal vector and $p$ the mean curvature, 
as before. This is precisely the IMCF condition. Notice, however, that IMCF in 
codimension one fixes uniquely the velocity of the flow while in codimension 
two, it still leaves room for imposing one extra scalar condition on the 
velocity. We shall use the name IMCF in both cases.

The requirement which could be used alternatively to remove (\ref{IP}) reads $ 
\Delta \psi - \Omega  = \alpha(\lambda)$ for some constant $\alpha(\lambda)$. 
By Gauss' law and by a trivial Fredholm argument, this  equation has a unique 
solution for $\psi$  iff $\alpha(\lambda) = \int_{\lambda} \Omega 
\bm{\eta_{S_{\lambda}}}$. Curiously, this integral is related to the Hawking  
mass itself, c.f.\ (\ref{Haw}). As for the IMCF condition, this one has a 
counterpart in codimension one as well, obtained by substituting the mean 
curvature $p^2$ for $H^2$ in $\Omega$, c.f.\ (\ref{Om}). This condition could 
be used in place of the IMCF to guarantee monotonicity of the Geroch mass. The 
question whether the corresponding flow exists will be discussed elsewhere. 

In the case where the $\vec{\xi}$ is allowed to be also null, the splitting 
(\ref{dMH2}) is not possible and we must return to (\ref{dMH1}). The last term 
in this expression can be made zero with sufficient generality only by 
imposing the IMCF condition $(\vec{\xi} \cdot \vec{H} )= \g(\lambda)$.

It only remains to consider the integral 
\begin{eqnarray}
I \equiv \int_{S_{\lambda}} 
\mbox{tr}_{S_{\lambda}} \left ( \vec{H} \cdot \nablanor \nablanor \vec{\xi} \right )
\bm{\eta_{S_{\lambda}}}.
\end{eqnarray}
\begin{lemma}
\label{LL}
Under the same hypotheses of Lemma \ref{derMH1}.
\begin{itemize}
\item[(a)] If $\vec{H} |_{S(\lambda)} = B \vec{\xi}^{\star} |_{S(\lambda)}$ 
for some function $B$ on the surface, then $I= 0$  provided  $\vec{\xi}$ is 
null all over $S_{\lambda}$ \underline{or} if $B$ is constant on $S_{\lambda}$. 
\item[(b)] Assume  $(\vec{\xi} \cdot \vec{H} ) = \g(\lambda)$ and $(\vec{\xi} 
\cdot \vec{H}^{\star} ) = c(\lambda)$ for some constant $c(\lambda)$ on 
$S_{\lambda}$. Then $I \ge 0$ provided (i) $\vec{\xi} \in C^{+}, \vec{H} \in 
C^{-}$, (ii) $\vec{\xi} \in C^{-}, \vec{H} \in C^{+}$, (iii) $\vec{\xi}, 
\vec{H} \in A^{+}$, \underline{or} (iv) $\vec{\xi}, \vec{H} \in A^{-}$ holds 
everywhere on $S_{\lambda}$
\end{itemize}
\end{lemma}
{\it Proof:}
Regarding part (a), if $\vec{\xi}$ is null on an open set $\U \subset \Slam$, 
we in fact have $\vec{H} = \pm B \vec{\xi}$ on $U$ and hence 
$\nablanor_{\vec{e}_A} \vec{\xi} = b_A \vec{\xi}$ on $U$, for some $b_A$. Thus, 
$\nablanor_{\vec{e}_B} \nablanor_{\vec{e}_A} \vec{\xi}$ is parallel to 
$\vec{\xi}$ and hence orthogonal to $\vec{H}$. Consequently $( \vec{H} \cdot 
\nablanor \nablanor \vec{\xi}) |_U=0$ holds irrespective of the values of $B$. 
If $\vec{\xi}$ is non-null everywhere then from the fact that $(\vec{\xi} 
\cdot \vec{H} ) = 0$ and $(\vec{\xi} \cdot \vec{H}^{\star} ) = \epsilon B 
e^{2\psi}$, (\ref{Derderxi}) implies that  the integrand in $I$ can be written 
as $- B \nablab_A \left (\xitwo U^A \right)$ which integrates to zero provided 
$B$ is constant. If $\vec{\xi}$ is non-null almost everywhere the same 
conclusion holds by continuity. Moreover, if $\vec{\xi}$ is null on an open 
set, the integral on that set is zero for any $B$, as shown before. Thus if 
$B$ is constant on $\Slam$ the integral vanishes irrespective of the causal 
character of $\vec{\xi}$.

Turning to (b), we first notice that if $\vec{\xi}$ is non-null everywhere, 
then (\ref{Derderxi}) together with the fact that $(\vec{\xi} \cdot \vec{H} )$ 
and $(\vec{\xi} \cdot \vec{H}^{\star} )$ are constants on $\Slam$ implies $I = 
\int_{\Slam} \Theta^L (\vec{\xi}) (\vec{H}^{\star}, \vec{\xi}^{\star} \, ) 
\bm{\eta_{S_{\lambda}}} \geq 0$, where Lemma \ref{ThetaL} has been used for 
the last inequality. To include the case when $\vec{\xi}$ may be null we use a 
continuity argument. The integral $I$ defines a map from $C^0(\Slam)\times 
C^2(\Slam) \rightarrow \mathbb{R}$, where the first factor refers to $\vec{H}$ 
and the second factor  to $\vec{\xi}$. This map is obviously continuous with 
respect to the supremum norm on these spaces. Thus, it is sufficient to 
observe that  the inequality holds for $\vec{H}, \vec{\xi} \in    \mbox{int} [ 
C_0^{+}(\Slam) \times C_2^{-}(\Slam)]  \cup \mbox{int} [ C_0^{-}(\Slam) \times 
C_2^{+}(\Slam)] \cup \mbox{int} [ A_0^{+}(\Slam) \times A_2^{+}(\Slam)]  \cup 
\mbox{int} [ A_0^{-}(\Slam) \times A_2^{-}(\Slam)]$ where $C_0^{+} (\Slam)$ 
denotes the set of vectors $\vec{h} \in C^0(\Slam)$ such that, for all $p \in 
\Slam$, $\vec{h} (p) \in C^{+}_p$. $C_2^{+} (\Slam)$ is defined analogously by 
replacing $C^0(\Slam)$ by $C^2(\Slam)$ and the other sets are defined 
similarly. Notice that this open set is dense in $[ C^{+}(\Slam) \times 
C^{-}(\Slam)]  \cup [ C^{-}(\Slam) \times C^{+}(\Slam)] \cup [ A^{+}(\Slam) 
\times A^{+}(\Slam)]  \cup [ A^{-}(\Slam) \times A^{-}(\Slam)]$, where we want 
to prove the Lemma. \hfill $\Box$

{\bf Remark 1}. Condition (a) holds for instance when 
$( \vec{H} \cdot \vec{\xi} )=0$ everwhere on $\Slam$ and $\vec{\xi}$
does not vanish on open sets of $\Slam$.



{\bf Remark 2.} This Lemma tells us that in the generic case, i.e. $(\vec{\xi} 
\cdot \vec{H})$ not identically zero,  we need to restrict the scalar product 
$(\vec{\xi} \cdot \vec{H}^{\star})$ to being constant on each leaf. We call 
this the {\it dual inverse mean curvature condition}. Such a condition is 
unnecessary in the time-symmetric case, where the Geroch-Hawking mass depends 
only on the intrinsic curvature of the hypersurface generated by the flow. 

Among the conditions (i) to (iv) that ensure monotonicity in part $(b)$ of 
Lemma \ref{LL}, only (iii) and (iv), i.e.\ velocity vector and mean curvature 
vector both achronal and to the same quadrant, give conditions which are 
compatible with those  ensuring non-negativity of (\ref{IG}) and $\Theta^T 
(\vec{H}^{\star},\vec{\xi}^{\star}\,)$. 

Regarding the case $\g(\lambda)=0$ on $S_{\lambda}$, the orthogonality of 
$\vec{H}$ and $\vec{\xi}$ implies that they cannot belong to the same 
quadrant, thus violating the non-negativity condition in Lemma \ref{LT}, 
unless $\vec{\xi}$ is null and $\vec{H} = B \vec{\xi}$, with a non-negative 
arbitrary {\it function} on $S_{\lambda}$. Thus, when the surface is 
marginally trapped and the flow is null, monotonicity is very easily achieved 
initially, the only condition being that the velocity is pointwise parallel to 
the mean curvature of the surface. Of course, the property of $\vec{H}$ being 
null will not be maintained by the flow and the flow vector will have to 
adjust, if at all possible, to keep a monotone flow.

Summarizing all the results above and noticing finally that the integral 
$\int_{\Slam} D.U (\vec{\xi} \cdot \vec{H}^{\star}) \bm{\eta_{S_{\lambda}}}$ 
in (\ref{dMH2}) vanishes provided the dual IMCF condition is satisfied, we can 
write down the following theorem which gives sufficient conditions on a 
spacetime flow of 2-surfaces in a spacetime so that the Hawking mass is 
monotone along the flow.

\begin{theorem}
\label{Th}
Let $S$ be an oriented, spacelike, closed, $C^2$ embedded 2-surface in a 
spacetime $(V,g)$. Let $\vec{\xi}$ be a $C^2$ normal vector field on
$S$, $\vec{H}$ the mean curvature vector of the surface and $M_H(S)$ its
Hawking mass. Let $S_{\lambda}$ be any flow of spacelike two-surfaces
starting at $S$ with $\vec{\xi}$ as normal component of the initial
velocity. Then, $d M_H (S_{\lambda})/d \lambda |_{\lambda=0} \geq 0$ holds 
whenever the following four conditions hold
\begin{itemize}
\item[(1)] The spacetime satisfies the dominant energy condition (\ref{ec}). 
\item[(2)] The mean curvature vector $\vec{H}$ is spacelike or null on $S$, 
and $\vec{\xi}$ points into the same causal quadrant as $\vec{H}$. \item[(3)] 
Either (a) the IMCF condition holds, i.e.\ $(\vec{\xi} \cdot \vec{H}) = \g_0$, 
where $\g_0$ is a non-negative constant, \underline{or} (b) $\xi^2 \neq 0$ 
everywhere on $S$ and 
\begin{equation}
\label{newHaw}
 \Delta \psi = \frac{1}{2} R(h_{\lambda}) -  \frac{1}{4} H^2 + \frac{1}{3} \Lambda - \alpha
\end{equation}
 where $\alpha$ is constant.
\item[(4)] The dual IMCF condition holds, i.e.\ $(\vec{\xi} \cdot 
\vec{H}^{\star} )  = \c_0$, where $\c_0$ is a constant. 
\end{itemize}
\end{theorem}

{\bf Remark.} We now assume that $\xi^2 \neq 0$ and that (3), (4) hold. Eqn.\ 
(\ref{dMH2}) can be reformulated so that the energy interpretation becomes 
transparent. We can write the change of $M_H$ as follows
\begin{eqnarray}
\frac{dM_{H}(S_{\lambda})}{d\lambda} = \int_{S_{\lambda}} 
\bm{\eta_{S_{\lambda}}} \left (\frac{1}{8\pi} (G_{\mu\nu} + \Lambda g_{\mu\nu}) + \Theta^T_{\mu\nu}
+ \Theta^L(\vec{\xi}\,)_{\mu\nu} \right ){\chi}^{\mu}{\xi}^{\star\nu}. 
\label{conservation}
\end{eqnarray}
Here we have introduced $\vec{\chi}=\vec{H}^\star\sqrt{|S_\lambda|/16\pi}$; 
the normalization is chosen such that $\vec{\chi}$ coincides with the 
asymptotically unit timelike Killing vector when $S$ is a 2-sphere on a 
time-symmetric slice in Schwarz\-schild. In more general situations with 
timelike Killing vectors, we are not aware if they necessarily coincide with 
$\vec{\chi}$. We note that on any marginally trapped surface, 
$\vec{H}^{\star}$ and therefore $\vec{\chi}$ is a null vector. A 
spin-coefficient version of the energy conservation law (\ref{conservation}) 
is given in \cite{Hay3a}. While $\Theta_T$ is a tensorial object (\ref{TTh}) 
depending only on the geometry, $\Theta_L$ depends on the velocity vector 
$\vec{\xi}$ as well. In any case, the components of these objects with respect 
to a null basis $(\vec{k}, \vec{l})$ with $(\vec{k} \cdot\vec{l}) = -\phi$ 
read 
\begin{eqnarray} 
8 \pi \Theta^T_{ll} & = & (\vec{l} \cdot \vec{\Pi}_{AB})^2 \qquad 8 \pi 
\Theta^T_{kk} = (\vec{k} \cdot \vec{\Pi}_{AB})^2 \qquad  
\Theta^T_{kl}=\Theta^T_{lk}=0. \nonumber\\ \Theta^L(\vec{\xi}\,)_{ll} \! & \! 
= \! &  \! \Theta^L(\vec{\xi}\,)_{kk}=0 \quad  8 \pi 
\Theta^L(\vec{\xi}\,)_{lk} = \phi (D\psi-U)^2 \quad 8 \pi 
\Theta^L(\vec{\xi}\,)_{kl} = \phi (D\psi+U)^2. \qquad 
\end{eqnarray}
These tensors may be interpreted respectively as encoding transverse and 
longitudinal modes of the gravitational field, with $\Theta \equiv \Theta^T + 
\Theta^L$ taking the same form as in a corresponding energy conservation law 
along black-hole horizons \cite{Hay3}. In particular, the components 
$\Theta^T_{ll}$ and $\Theta^T_{kk}$ can be interpreted as energy densities of 
ingoing and outgoing gravitational radiation, taking the same form as standard 
definitions at null infinity in an asymptotically flat spacetime. In that 
case, the conservation law corresponding to (\ref{conservation}) is the Bondi 
mass equation \cite{Hay4}. 



We proceed by discussing the requirements of Theorem \ref{Th} in some detail, 
restricting ourselves to the IMCF condition, i.e.\ $(3.a)$. 

First a remark on terminology: recalling that  $\pounds_{\vec{\xi}} \, \etaS = 
(\vec{\xi} \cdot \vec{H})\etaS $, we see that  the first of these conditions 
implies that the area element is preserved under the flow $\vec{\xi}$, while 
the second one implies the same along the dual $\vec{\xi}^{\star}$. This 
motivates the terminology {\it uniformly expanding flow} (UEF) \cite{Hay1} for 
both $(\vec{\xi} \cdot \vec{H}) = \g (\lambda)$ and $(\vec{\xi} \cdot 
\vec{H}^{\star}) = c(\lambda)$, while we reserve the term {\it inverse mean 
curvature flow} (IMCF) for the first one alone. The UEFs have some interesting 
special cases which we discuss in turn.

We first assume that the flow velocity is null everywhere, while the mean 
curvature vector of the surface is achronal everywhere. To obtain 
monotonicity, we impose the IMCF condition with $\g(\lambda)=1$. This implies 
the dual IMCF condition since $\vec{\xi}^{\star} = \pm \vec{\xi}$, and 
therefore
\begin{eqnarray*}
\left ( \vec{\xi} \cdot \vec{H}^{\star} \right)  = - 
\left ( \vec{\xi}^{\star} \cdot \vec{H} \right) 
= \mp \left ( \vec{\xi} \cdot \vec{H} \right ) = \mp 1.
\end{eqnarray*}
All the conditions of the theorem hold provided $\vec{\xi}$ and $\vec{H}$ 
point into the same quadrant. Notice that the IMCF condition actually prevents 
$\vec{\xi}$ or $\vec{H}$ from vanishing anywhere on the surface. This null 
monotone flow was discovered by Hayward \cite{Hay1}. Since $\vec{\xi}$ is null 
and hypersurface orthogonal, the flow is geodesic, i.e. $\nabla_{\vec{\xi}} \, 
\vec{\xi}= Y \vec{\xi}$ for some function $Y$. In terms of an affinely 
parametrized geodesic with tangent $l$, we have $\xi = \theta_{\vec{l}}^{-1} 
\vec{l}$ and the flow can be determined by solving the ODE $\xi (\lambda) = 
\theta_{\vec{l}}^{-1} \vec{l} (\lambda) = 1$. We emphasize that this is the 
only case for which we obtain local existence.

We next assume that $\vec{H}$ is spacelike everywhere, i.e.\ $H^2 > 0$, which 
allows us to write the velocity vector in the form
\begin{eqnarray}
\vec{\xi} = H^{-2} \left ( \vec{H} - c(\lambda) \vec{H}^{\star} \right ),
\label{Flow}
\end{eqnarray}
choosing $\g(\lambda)=1$. A special case is that $c(\lambda)$ vanishes 
identically, so that the velocity vector of the flow is $\vec{\xi} = 
H^{-2}\vec{H}$. This flow was mentioned by Huisken and Ilmanen in \cite{HI} 
and analyzed in more detail by Frauendiener \cite{JF}, who showed monotonicity 
in an explicit way. In the general case covered by (\ref{Flow}), we can set 
$a(\lambda) = 1$ so that the velocity flow contains one arbitrary function of 
one variable $\c(\lambda)$ satisfying $| \c(\lambda) | \leq  1$. In the 
particular case $|\c(\lambda)|<1$, this is the spacetime version of the flows 
found by Malec, Mars and Simon \cite{MMS} in terms of initial data sets. 
Indeed, in this case the vector $\vec{\xi}$ in (\ref{Flow}) is spacelike by 
construction, so that the hypersurface $\Sigma$ generated by the flow of 
2-surfaces will be spacelike, if it exists. 
Thus, when the flow is viewed as a flow of 2-surfaces in this 3-dimensional 
Riemannian manifold, the IMCF condition $(\vec{\xi} \cdot \vec{H}) = 
\g(\lambda)$  translates into the condition that $S$ flows within $\Sigma$ as 
an inverse mean curvature flow. The other condition $ ( \vec{\xi} \cdot 
\vec{H}^{\star} ) = \c(\lambda)$ becomes a condition involving $p$, the mean 
curvature of $S \subset  \Sigma$, and $q$, the trace on $S$ of the extrinsic 
curvature of $\Sigma$ as a hypersurface in spacetime, which explicitly reads 
$q/p = \c(\lambda)$. This is precisely the condition obtained in \cite{MMS} 
(see also \cite{MMS2}) in order to ensure monotonicity of the Hawking mass in 
an initial data setting. These flows  significantly generalize flows along the 
inverse mean curvature vector, since the one-parameter freedom after fixing 
$a(\lambda) = 1$ allows flows from a given initial surface to cover a 
spacetime region. In particular, while $\vec{H}/|H|^2$ can only approach 
spacelike infinity in an asymptotically flat spacetime, $\vec{\xi}$ of the 
form (\ref{Flow}) can also approach null infinity. This suggests the 
possibility to prove the Penrose inequality (\ref{PI}) not only for the ADM 
mass $M$ but also for the Bondi mass $M_B\le M$. The latter inequality  is 
significantly stronger than the former, but it also follows from the heuristic 
argument involving cosmic censorship \cite{Pen}.

The final special case of the UEF guaranteeing monotonicity of the Hawking 
mass are surfaces with null $\vec{H}$ and velocity vector which is either 
spacelike or null, towards the same quadrant as $\vec{H}$. Since $H = \pm 
H^{\star}$, the dual IMCF condition implies the IMCF condition, as in the case 
where $\vec{\xi}$ is null. The surface $S$ is restricted to being a future 
(past) marginally trapped surface, which is precisely the starting surface for 
flows aiming at proving the Penrose inequality. Given $\vec{H}$, the velocity 
vector $\vec{\xi}$ at $S$ has a freedom of one arbitrary {\it function}. Thus, 
we are allowed to flow out of a marginally trapped surface along a large class 
of flows for which $M_H$ is monotone.

\section{On existence of uniformly expanding flows}

In this section we discuss the prospects of obtaining existence of the flows 
defined above. Recall that the method of Huisken and Ilmanen in codimension 1 
rests mainly on two properties of the IMCF: firstly, the embedding functions 
determining the flow satisfy a parabolic system. Secondly, the level sets of 
the flow satisfy a degenerate elliptic system. As to codimension 2, Huisken 
and Ilmanen noticed that the embedding functions describing a flow along the 
inverse mean curvature vector, i.e.\ (\ref{Flow}) with $c \equiv 0$, satisfy a 
backward-forward parabolic system. There is no general theory which would 
guarantee local existence of solutions to such systems, and counter-examples 
indicate that such systems have to be treated with care. Unfortunately, as we 
will see below, the system keeps this vicious behaviour for any choice of $c$.

Nevertheless, the more important component on the work of Huisken and
Ilmanen is the weak (variational) formulation of the level set formulation
for the flow. We have obtained a generalization of this formulation
which we will describe in turn.

We first give the description in terms of embedding functions 
 $\Phi^{\alpha}(\lambda, x^A)$ for a 2-surface $S$
embedded in spacetime, following the  notation introduced in Sect.\ 2. In 
local coordinates we have $\xi^{\mu} = \partial \Phi^{\mu}/\partial \lambda$ 
and the tangent vectors to $S$ can be written as $e^{~~\mu}_{A} = \partial 
\Phi^{\mu}/\partial x^A$. For the mean curvature vector $H^{\mu}$ we obtain 
\begin{eqnarray}
H^{\mu} = h^{AB}  K^{\mu}_{AB} =   
 - \Delta \Phi^{\mu} - h^{AB}  \left. \Gamma^{\mu}_{~\nu\rho}
\right|_{x=\Phi^{\alpha}} 
\frac{\partial \Phi^{\nu}}{\partial \xi^A} \frac{\partial \Phi^{\rho}}{\partial
\xi^B},
\label{emb}
\end{eqnarray}
where $\Gamma^{\mu}_{~\nu\rho}$ are the  Christoffel symbols of the ambient
space, and $\Delta$ is the Laplacian on $S$.
In this way, (\ref{Flow}) becomes a system of PDEs of the form
\begin{eqnarray}
\frac{\partial \Phi^{\alpha}}{\partial \lambda} = 
J^{\alpha}\left(\lambda,x^A, \Phi^{\beta}, \frac{\partial \Phi^{\beta}}{\partial x^A}, 
\frac{\partial^2 \Phi^{\beta}}{\partial x^A \partial x^B} \right).
\label{PDE}
\end{eqnarray}
To determine the type of this system we need to calculate the eigenvalues of
the symmetrized part $Q + Q^{\dagger}$ of the matrix
\begin{equation}
Q^{\alpha}_{~\beta} \equiv \frac{\partial J^{\alpha}}
{\partial \left(\frac{\partial^2 \Phi^{\beta}}{\partial x^A \partial x^B} \right)} z^A z^B
\end{equation}
where $z^A$ is a unit vector in $ T^{\Vert}_p$, and $\dagger$ denotes the 
transpose with respect to some positive definite metric.
Restricting ourselves now to  codimension two, we obtain for the linearization 
of the operators $H$ and $H^{\star}$ acting on $\Phi$,
\begin{equation}
\frac{\partial H^{\alpha}}{\partial \left(\frac{\partial^2 \Phi^{\beta}}{\partial x^A \partial x^B}
\right)} z^A z^B =  -  P^{~\alpha}_{\bot~\beta}
\qquad
\frac{\partial H^{\star \alpha}}{\partial \left(\frac{\partial^2 \Phi^{\beta}}{\partial x^A \partial x^B}
\right)} z^A z^B = - \eta_{S ~\beta}^{\star \alpha}
\end{equation}
where we have employed the dualized 2-form $\bm{\eta_S^{\star}}$:
\begin{eqnarray*}
\eta_{S ~\beta}^{\star \alpha} = \frac{1}{2} \eta^{\alpha}_{\,\,\,\beta\gamma\delta}
 e^{\gamma}_A e^{\delta}_B \eta_{S}^{AB},
\end{eqnarray*}
It follows that 
\begin{equation}
Q^{\alpha}_{~\beta} = \frac{1}{H^2} \left[\left(\frac{2
H^{\alpha}H_{\beta}}{H^2} - P^{\alpha}_{\beta} \right) - 
\frac{c(\lambda)}{H^2} \left( H^{\alpha}H^{\star}_{\beta} + H^{\star 
\alpha}H_{\beta} \right) \right].
\end{equation}
To do the symmetrization, we can choose the metric $L_{\alpha \beta} = H^{-2} 
(H_{\alpha} H_{\beta} + H_{\alpha}^{\star} H_{\beta}^{\star})$ so that the 
symmetrized part of $Q^{\alpha}_{\beta}$ is independent of $c(\lambda)$ and 
reads
\begin{equation}
\left( Q + Q^{\dagger}\right)^{\alpha}_{~\beta} = 
\frac{2}{H^4} \left( H^{\alpha}H_{\beta} + H^{\star \alpha}H^{\star}_{\beta} \right)
\end{equation}
This matrix has $H^{\alpha}$ and $H^{\star \alpha}$ as eigenvectors, 
with eigenvalues $2/H^2$ and $2/H^{\star 2} = -2/H^2$, respectively.
Therefore $\left( Q + Q^{\dagger}\right)^{\alpha}_{~\beta}$ is indefinite and 
(\ref{Flow}) is a so-called forward-backward parabolic system. 
The previous reasoning applies in codimension one if all $H^{\star}$ are
removed, and yields parabolicity, as well known.
It can be easily checked that the reasoning above
is independent of the metric $L_{\mu\nu}$ used for symmetrization.

Turning now to the level set formulation, we start with recalling the case of 
codimension 1. One can envision the family of surfaces $S(\lambda)$ as level 
sets of a level set function $u(x)$ defined on a spacelike hypersurface 
$\Sigma$ of the spacetime. If $u$ is twice differentiable, the inverse mean 
curvature flow condition can be written in terms of $u$ as a degenerate 
elliptic equation, namely

\begin{equation}
\label{degel}
{\cal D}. \left(\frac{{\cal D} u}{|{\cal D} u|} \right) = |{\cal D} u|
\end{equation}
where ${\cal D}$ is the covariant derivative on $\Sigma$, and ${\cal D}.W$  
denotes the divergence of some vector $W$. The variational formulation reduces 
the requirements on differentiability. While (\ref{degel}) are not 
Euler-Lagrange equations of a functional of $u$ as the only variable, they can 
nevertheless be obtained by minimizing the functional 
\begin{equation}
\label{funct1} E^K_u(v) = \int_{\Sigma \cap K}\left( |{\cal D} v| + v |{\cal 
D} u|\right)
\end{equation}
with respect to variations of $v$ on any compact set $K$. One can imagine this 
variational principle as a two-step procedure in which the term $|{\cal D} u|$ 
on the right hand side of (\ref{degel}) is first fixed or ``frozen'' while 
$E_{u}$ is varied with respect to $v$, and in the second step $|{\cal D} u|$ 
is put equal to the $|{\cal D} v|$ with the function $v$ obtained from the 
minimization, with derivatives understood in a weak sense.

As to codimension 2, it would be desirable to have some generalization of 
(\ref{degel}) with an extension of $u$ to the extra dimension. While from the  
equation itself it is not clear how this could be accomplished, we turn 
directly to the variational formulation, which does, in fact, involve a 
natural extension of $u$.

We start from some initial two-surface $S$ embedded in some three-surface 
$\Sigma$. The latter is arbitrary at this stage, but will be fixed 
automatically by the variation principle. We consider the level sets 
$S(\lambda) \subset \Sigma$ given by the level set function $u(\lambda)$ of 
the IMCF near $S$, obtained via the Huisken-Ilmanen procedure.  We denote by 
$\nu (\lambda)$ and $\mu (\lambda) = \nu^{\star}(\lambda)$ the unit normals to 
$S(\lambda)$ in the future timelike direction normal to $\Sigma$ and in the 
spacelike direction tangent to $\Sigma$, respectively, assuming their 
existence. The variational principle should now move the flow in the direction 
of $\vec{\xi}$ given by (\ref{Flow}), i.e.\ $\Sigma$ should have $\vec{\xi}$ 
as tangent vector whenever the latter is defined. Equivalently, the mean 
curvature vector to $S_{\lambda}$ should be orthogonal to $\nu + c(u)\mu$, and 
this latter property is used in the weak formulation. We extend the function 
$u$ off $\Sigma$ in some neighbourhood of $\Sigma$ in such a way that the 
directional derivative of $u$ in the direction $\nu + c(u)\mu$, whose 
existence we may assume, is zero everywhere on $S_{\lambda}$. We denote by 
$\Upsilon$ any other spacelike hypersurface in this neighborhood  which is 
equal to $\Sigma$ outside some compact region $\Gamma$, and we take $v$ to be 
a real-valued function on $\Upsilon$ which agrees with $u$ outside of $\Gamma$ 
as well. We then define the functional
\begin{equation}
E^{\Gamma}_{\Sigma, u, c}(\Upsilon, v) = \int_{\Upsilon \cap \Gamma} 
\left(|{\cal D} v| + v |{\cal D} u|\right). 
\end{equation}
In analogy with (\ref{funct1}), the ``frozen'' structure now consists of the
hypersurface $\Sigma$ together with the level sets $u$, extended as above to a 
neighbourhood of $\Sigma$. The objects which are varied  are $\Upsilon$ and 
$v$, and in the second step, $\Sigma$ and $u$ are adjusted according to the 
results of the variation. This justifies the following.

\begin{definition}
Given $c(\lambda) \in [-1,1]$, then 
$(\Sigma,u)$ is a weak solution to the corresponding uniformly expanding
flow in an open region $\Omega$ if it is a critical point of the functional
$E^{\Gamma}_{\Sigma, u, c}(\cdot,\cdot)$ for all compact $\Gamma \subset \Omega$.
\end{definition}

We will also say that the 2-surfaces $u$ are weak solutions to equation 
(\ref{Flow}).  Since $u(x)$ may have constant regions, this allows the family 
of surfaces defined by the level sets of $u(x)$ to jump, just as in the 
Huisken-Ilmanen case.

Given this weak definition, we may check that the above weak solutions agree 
with the usual smooth solutions when they exist. We first use the fact that 
$(\Sigma,u)$ is a critical point of $E^K_{\Sigma, u, c}(\cdot,\cdot)$ with 
respect to variations of $\Sigma$.  Consider a variation $\Upsilon$ of 
$\Sigma$ which is a compactly supported bump function on $\Sigma$ times $\nu + 
c(u)\mu$.  Note that $u$ does not change to first order in these directions.  
Hence, to first order, it is still true that 
\[
E = \int_{\Upsilon \cap \Gamma} \left(|{\cal D} u| + u |{\cal D} u|\right),
\]
where we may extend $u$ to be constant in this variation since we are at a 
critical point. But by the co-area formula, if we choose $\Gamma$ to be the 
region where $a \le u(x) \le b$, 
\[
E = \int_a^b (1+\lambda)|S(\lambda)| d\lambda,
\]
where $|S(\lambda)|$ denotes that area of $S(\lambda)$. Since we are at a 
critical point, the first variation of $E$ must be zero, which, in the case 
that $S(\lambda)$ is a smooth family of surfaces, follows if and only if our 
variation direction is orthogonal to the mean curvature vector of 
$S(\lambda)$.  Equivalently, it follows that $\vec\xi_c$ from equation 
(\ref{Flow}) must be {\it tangent} to $\Sigma$.

What remains to be done is to check that the actual flow vector, which is also 
tangent to $\Sigma$ and orthogonal to each $S(\lambda)$, has the same length 
as $\vec\xi_c$.  Again we use the fact that $(\Sigma,u)$ is a critical point 
of $E^{\Gamma}_{\Sigma, u, c}(\cdot,\cdot)$ with respect to variations of 
$u(x)$.  This implies that $u(x)$ is a weak solution to inverse mean curvature 
flow in $\Sigma$ in the sense defined by Huisken and Ilmanen.  Thus, the level 
sets $S(\lambda)$ of $u(x)$ have the uniformly expanding property that the 
rate of change of the area form of the surfaces equals the area form itself.  
This proves that the actual flow vector of the surfaces is not only in the 
same direction as $\vec\xi_c$ but also has the same length, and therefore they 
are equal.  Thus, the smooth flow case agrees with this weak definition.

To conclude this exposition we wish to emphasize that  having a definition of 
a weak solution to equation (\ref{Flow}) is still very far from an existence 
theorem.  However, the correct notion of a weak solution is a prerequisite to 
an existence theorem along the lines of Huisken and Ilmanen.

\bigskip{\bf Acknowledgements.} 
M.M. and W.S. are grateful to Tom Ilmanen for helpful discussions. S.A.H. was 
supported by NSF grants PHY-0090091, PHY-0354932 and the Eberly research funds 
of Penn State. M.M. and W.S. were supported by the projects BFME2003-02121 of 
the Spanish Ministerio de Educaci\'on y Tecnolog\'{\i}a and SA010CO of the 
Junta de Castilla y Le\'on. W.S. was also supported in part by FWF (Austria), 
grant P14621-N05.

\end{document}